\documentclass[pra,aps,twocolumn,showpacs]{revtex4}

\usepackage{amsmath}
\usepackage{bm}
\usepackage{graphicx}

\begin{document}

\title{Kibble-Zurek mechanism in a quenched ferromagnetic
Bose-Einstein condensate
}

\author{Hiroki Saito$^1$}
\author{Yuki Kawaguchi$^2$}
\author{Masahito Ueda$^{2,3}$}
\affiliation{$^1$Department of Applied Physics and Chemistry, The
University of Electro-Communications, Tokyo 182-8585, Japan \\
$^2$Department of Physics, Tokyo Institute of Technology,
Tokyo 152-8551, Japan \\
$^3$Macroscopic Quantum Control Project, ERATO, JST, Bunkyo-ku, Tokyo
113-8656, Japan
}

\date{\today}

\begin{abstract}
The spin vortices are shown to be created through the Kibble-Zurek (KZ)
mechanism in a quantum phase transition of a spin-1 ferromagnetic
Bose-Einstein condensate, when the applied magnetic field is quenched
below a critical value.
It is shown that the magnetic correlation functions have finite
correlation lengths, and magnetizations at widely separated positions grow
in random directions, resulting in spin vortices.
We numerically confirm the scaling law that the winding number of spin
vortices is proportional to the square root of the length of the closed
path, and for slow quench, proportional to $\tau_{\rm Q}^{-1/6}$ with
$\tau_{\rm Q}$ being the quench time.
The relation between the spin conservation and the KZ mechanism is
discussed.
\end{abstract}


\maketitle

\section{Introduction}

Spontaneous symmetry breaking in a phase transition
produces local domains of an order parameter.
If domains are separated by such a long distance that they cannot exchange
information, local domains grow initially with random phases and 
eventually give rise to topological defects when they overlap.
This scenario of topological-defect formation in continuous-symmetry
breaking is known as the Kibble-Zurek (KZ) mechanism~\cite{Kibble,Zurek},
which originally predicted the cosmic-string and monopole formation in
the early Universe~\cite{Kibble}, and has since been applied to a wide
variety of systems.
Experimentally, the KZ mechanism has been examined in liquid
crystals~\cite{Chuang,Bowick}, superfluid $^4{\rm He}$~\cite{Hendry} and
$^3{\rm He}$~\cite{Ruutu,Bauerle}, an optical Kerr medium~\cite{Ducci},
Josephson junctions~\cite{Monaco,Carmi}, and superconducting
films~\cite{Maniv}.

Recently, spontaneous magnetization in a spinor Bose-Einstein condensate
(BEC) has attracted much interest as a new system for studying the KZ
mechanism~\cite{Sadler,Saito07,Lamacraft,Uhlmann}.
In the experiment performed by the Berkeley group~\cite{Sadler}, a BEC of
$F = 1$ $^{87}{\rm Rb}$ atoms are prepared in the $m = 0$ state, where
$F$ is the hyperfine spin and $m$ is its projection on the direction of
the magnetic field.
By quench of the magnetic field, say in the $z$ direction, magnetization
appears in the $x$-$y$ plane.
Since the spinor Hamiltonian is axisymmetric with respect to the $z$ axis,
the magnetization in the $x$-$y$ direction breaks the U(1) symmetry in the
spin space.
Thus, local domain formation is expected to lead to topological defects
--- spin vortices --- through the KZ mechanism.

However, the origin of the spin vortices observed after the quench in the
Berkeley experiment~\cite{Sadler} cannot be attributed to the KZ
mechanism.
In fact, in Ref.~\cite{Sadler}, the spin correlation extends over the
entire system (at least in the $x$ direction) and the domains are not
independent with each other.
We have shown in Ref.~\cite{Saito07} that the origin of the observed spin
vortices is initial spin correlation due to the residual $m = \pm 1$
atoms, which forms domain structure followed by spin vortex
creation~\cite{Saito06}.
In order to realize the KZ mechanism in this system, i.e., in order to
ensure that the magnetic domains grow independently, the size of the
system must be much larger than the spin correlation length and the
long-range correlation in the initial spin state must be absent.
The aim of the present paper is to show that under these conditions spin
vortices are generated through the KZ mechanism.

In the present paper we will consider 1D-ring and 2D-disk geometries.
We will show that in the 1D ring the average spin winding number after the
quench is proportional to the square root of the system size, which is in
agreement with the KZ prediction~\cite{Zurek}.
In 2D the winding number along a path with radius $R$ is also proportional
to $R^{1/2}$ as long as $R$ is much larger than the vortex spacing, while
it is proportional to $R$ for small $R$.
When the magnetic field is quenched slowly, the winding number is shown to
be proportional to $\tau_{\rm Q}^{-1/6}$ with $\tau_{\rm Q}$ being the
quench time.
This power law can be understood by Zurek's simple
discussion~\cite{Zurek}.

The spinor BEC is different from the other systems in which the KZ
mechanism has been observed, in that the total spin is conserved when the
quadratic Zeeman energy $q$ is negligible.
This fact is seemingly incompatible with the KZ postulate, since the
magnetic domains must be correlated with each other so that the total
magnetization vanishes.
We will show that for $q = 0$ small magnetic domains are aligned to cancel
out the local spin averaged over the correlation length, and that they are
independent with each other over a greater length scale; the spin
conservation is thus compatible with the KZ mechanism.

The present paper is organized as follows.
Section~\ref{s:bogo} analyzes spontaneous magnetization of a spin-1 BEC
and the resultant magnetic correlation functions using the Bogoliubov
approximation.
Section~\ref{s:num} performs numerical simulations of the dynamics of
quenched BECs in 1D and 2D, and shows that the KZ mechanism does emerge in
the present system.
Section~\ref{s:conc} provides conclusions.

\section{Bogoliubov analysis of a quenched ferromagnetic Bose-Einstein
condensate}
\label{s:bogo}

\subsection{Hamiltonian for the spin-1 atoms}

We consider spin-1 bosonic atoms with mass $M$ confined in a potential
$V_{\rm trap}(\bm{r})$.
The noninteracting part of the Hamiltonian is given by
\begin{equation} \label{H0}
\hat H_0 = \int d\bm{r} \sum_{m = -1}^1 \hat\psi_m^\dagger(\bm{r}) \left[
-\frac{\hbar^2}{2M} \bm{\nabla}^2 + V_{\rm trap}(\bm{r}) \right]
\hat\psi_m(\bm{r}),
\end{equation}
where $\hat\psi_m(\bm{r})$ annihilates an atom in magnetic sublevel $m$ of
spin at a position $\bm{r}$.

The interaction between atoms with $s$-wave scattering is described by
\begin{equation} \label{Hint}
\hat H_{\rm int} = \frac{1}{2} \int d\bm{r} :\left[ c_0 \hat\rho^2(\bm{r})
+ c_1 \hat{\bm{F}}^2(\bm{r}) \right]:,
\end{equation}
where the symbol $::$ denotes the normal order and
\begin{eqnarray}
\label{rho}
\hat\rho(\bm{r}) & = & \sum_{m = -1}^1 \hat\psi_m^\dagger(\bm{r})
\hat\psi_m(\bm{r}), \\
\label{F}
\hat{\bm{F}}(\bm{r}) & = & \sum_{m, m'} \hat\psi_m^\dagger(\bm{r})
\bm{f}_{mm'} \hat\psi_{m'}(\bm{r}),
\end{eqnarray}
with $\bm{f} = (f_x, f_y, f_z)$ being the spin-1 matrices.
The interaction coefficients in Eq.~(\ref{Hint}) are given by
\begin{subequations}
\begin{eqnarray}
c_0 & = & \frac{4 \pi \hbar^2}{M} \frac{a_0 + 2 a_2}{3}, \\
c_1 & = & \frac{4 \pi \hbar^2}{M} \frac{a_2 - a_0}{3},
\end{eqnarray}
\end{subequations}
where $a_S$ is the $s$-wave scattering lengths for two colliding atoms
with total spin $S$.

When magnetic field $\bm{B}$ is applied, the linear Zeeman effect rotates
the spin around the direction of $\bm{B}$ at the Larmor frequency.
Since $\hat H_0$ and $\hat H_{\rm int}$ are spin-rotation invariant and we
assume the uniform magnetic field, the linear Zeeman term has only a
trivial effect on spin dynamics --- uniform rotation of spins about
$\bm{B}$ --- which is therefore ignored.
The quadratic Zeeman effects for an $F = 1$ $^{87}{\rm Rb}$ atom is
described by
\begin{equation} \label{Hq}
\hat H_q = \frac{\mu_{\rm B}^2}{4 E_{\rm hf}} \int d\bm{r} \sum_{m, m'}
\hat\psi_m^\dagger(\bm{r}) \left[ (\bm{B} \cdot \bm{f})^2 \right]_{mm'}
\hat\psi_{m'}(\bm{r}),
\end{equation}
where $\mu_{\rm B}$ is the Bohr magneton and $E_{\rm hf} > 0$ is the
hyperfine splitting energy between $F = 1$ and $F = 2$.
The total Hamiltonian is given by the sum of Eqs.~(\ref{H0}),
(\ref{Hint}), and (\ref{Hq}),
\begin{equation}
\hat H = \hat H_0 + \hat H_q + \hat H_{\rm int}.
\end{equation}

\subsection{Time evolution in the Bogoliubov approximation}

In the initial state, all atoms are prepared in the $m = 0$ state.
We study the spin dynamics of the system using the Bogoliubov
approximation with respect to this initial state.
For simplicity, we assume $V_{\rm trap} = 0$ in this section.

In the Bogoliubov approximation, the BEC part in the field operator is
replaced by a c-number.
In the present case, we write the $m = 0$ component of the field operator
as
\begin{equation}
\hat\psi_0(\bm{r}) = e^{-i c_0 n t / \hbar} \left[ \sqrt{n} +
\delta \hat\psi_0(\bm{r}) \right],
\end{equation}
where $n$ is the atomic density.
We expand $\hat\psi_{\pm 1}(\bm{r})$ as
\begin{equation}
\hat\psi_{\pm 1}(\bm{r}) = e^{-i c_0 n t / \hbar} \sum_{\bm{k}}
\frac{1}{\sqrt{V}} e^{i \bm{k} \cdot \bm{r}} \hat a_{\pm 1, \bm{k}},
\end{equation}
where $V$ is the volume of the system and $\hat a_{\pm 1, \bm{k}}$ is the
annihilation operator of an atom in the $m = \pm 1$ state with wave vector
$\bm{k}$.
Keeping only up to the second order of $\delta \hat\psi_0(\bm{r})$ and
$\hat\psi_{\pm 1}(\bm{r})$ in the Hamiltonian, we obtain the Heisenberg
equation of motion for $\hat a_{\pm 1, \bm{k}}$ as
\begin{equation} \label{dadt}
i \hbar \frac{d \hat a_{\pm 1, \bm{k}}(t)}{dt} = (\varepsilon_k + q + c_1
n) \hat a_{\pm 1, \bm{k}}(t) + c_1 n \hat a_{\mp 1, -\bm{k}}^\dagger(t),
\end{equation}
where $\varepsilon_k = \hbar^2 k^2 / (2M)$ and $q = \mu_{\rm B}^2 B^2 / (4
E_{\rm hf})$.
The magnetic field is assumed to be applied in the $z$ direction.
The solution of Eq.~(\ref{dadt}) is obtained as
\begin{eqnarray} \label{at}
\hat a_{\pm 1, \bm{k}}(t) & = & \left( \cos \frac{E_k t}{\hbar} - i
\frac{\varepsilon_k + q + c_1 n}{E_k} \sin \frac{E_k t}{\hbar} \right)
\hat a_{\pm 1, \bm{k}}(0) \nonumber \\
& & - \left( i \frac{c_1 n}{E_k} \sin
\frac{E_k t}{\hbar} \right) \hat a_{\mp 1, -\bm{k}}^\dagger(0),
\end{eqnarray}
where
\begin{equation}
E_k = \sqrt{(\varepsilon_k + q) (\varepsilon_k + q + 2 c_1 n)}.
\end{equation}

When $E_k$ is imaginary, the corresponding modes are dynamically unstable
and grow exponentially.
Since $c_1 < 0$ and $q > 0$ for $F = 1$ $^{87}{\rm Rb}$ atoms, the
exponential growth occurs for
\begin{equation}
q < 2 |c_1| n \equiv q_{\rm c}.
\end{equation}
This critical value of $q$ agrees with the phase boundary between the
polar phase and the broken-axisymmetry phase~\cite{Stenger,Murata}.
When $q \leq q_{\rm c} / 2$, the wave number of the most unstable mode is
\begin{equation} \label{kmu}
k_{\rm mu} = \pm \sqrt{\frac{2M}{\hbar^2} \left( \frac{q_{\rm c}}{2} - q
\right)},
\end{equation}
and when $q_{\rm c} / 2 < q < q_{\rm c}$, $k_{\rm mu} = 0$.

\subsection{Fast quench}

We consider the situation in which $q$ is much larger than the other
characteristic energies for $t < 0$, and $q$ is suddenly quenched below
$q_{\rm c}$ at $t = 0$.
During $t < 0$, the time evolution in Eq.~(\ref{at}) is $\hat a_{\pm 1,
\bm{k}}(t) \simeq e^{-i q t / \hbar} \hat a_{\pm 1, \bm{k}}(0)$, and the $m =
\pm 1$ state remains in the vacuum state.
For $t > 0$, we obtain the time evolution of the density of the $m
= \pm 1$ component as
\begin{eqnarray} \label{density}
\left\langle \hat\psi_{\pm 1}^\dagger(\bm{r}, t) \hat\psi_{\pm 1}(\bm{r},
t) \right\rangle & = & \frac{1}{V} \sum_{\bm k} \left| \frac{c_1 n}{E_k}
\sin \frac{E_k t}{\hbar} \right|^2
\nonumber \\
& \simeq & \frac{1}{V} \sum_{k < k_{\rm c}} \frac{q_{\rm c}^2}{16 |E_k|^2}
e^{2 |E_k| t / \hbar},
\end{eqnarray}
where the expectation value is taken with respect to the vacuum state of
the $m = \pm 1$ component.
In the second line of Eq.~(\ref{density}), we have kept the unstable modes
alone with $k < k_{\rm c} \equiv \sqrt{2M (q_{\rm c} - q)} / \hbar$ by
assuming that $|E_k| t / \hbar \gg 1$.
This result indicates that the $m = \pm 1$ components grow exponentially
after the quench.

Since the operator $\hat\psi_0$ in Eq.~(\ref{F}) is replaced by $\sqrt{n}$
in the Bogoliubov approximation, the magnetization operator $\hat F_+ =
\hat F_-^\dagger = \hat F_x + i \hat F_y$ has the form,
\begin{equation}
\hat F_+(\bm{r}) = \sqrt{2 n} \left[ \hat \psi_1^\dagger(\bm{r}) + \hat
\psi_{-1}(\bm{r}) \right].
\end{equation}
Using Eq.~(\ref{at}), the time evolution of the correlation function is
calculated to be
\begin{subequations}
\begin{eqnarray}
& & \left\langle \hat F_+({\bm r}, t) \hat F_-({\bm r}', t) \right\rangle
\nonumber \\
& = & \frac{2n}{V} \sum_{\bm{k}} \left| \cos \frac{E_k t}{\hbar} + i
\frac{\varepsilon_k + q}{E_k} \sin \frac{E_k t}{\hbar} \right|^2
e^{i \bm{k} \cdot (\bm{r} - \bm{r}')}
\label{ffa}
\nonumber \\
\\
\label{ff}
& \simeq & \frac{n}{2V} \sum_{k < k_{\rm c}} \frac{q_{\rm
c}}{q_{\rm c} - q - \varepsilon_k} e^{2 |E_k| t / \hbar + i \bm{k} \cdot
(\bm{r} - \bm{r}')},
\end{eqnarray}
\end{subequations}
where in the second line we have kept the unstable modes alone.

From the exponential factor in Eq.~(\ref{ff}), we see that the sum is
contributed mostly from $\bm{k}$ around the mode with maximum $|E_k|$.
The denominator in the summand of Eq.~(\ref{ff}) is much smoother than the
exponential factor if $q$ is not close to $q_{\rm c}$, and then we
approximate $\varepsilon_k$ with $\varepsilon_{\rm mu} = \hbar^2 k_{\rm
mu}^2 / (2M)$ in the denominator.
We expand $2 |E_k| t / \hbar$ around $k_{\rm mu}$ in the exponent as
\begin{equation} \label{Ekexpand}
\frac{2 |E_k| t}{\hbar} = \frac{t}{\tau} \left( 1 - \frac{1}{4} \xi^2
\Delta k^2 - \frac{1}{256} \Xi^4 \Delta k^4 \right) + O(\Delta k^6),
\end{equation}
where $\Delta k = k - k_{\rm mu}$.
It is clear that $\tau$ sets the time scale for the exponential growth.
The magnetization is observed when it sufficiently grows, i.e., $t \sim
\tau$.
Replacing the summation with the Gaussian integral in Eq.~(\ref{ff}), we
find that $\xi$ represents the correlation length.
For $q < q_{\rm c} / 2$, $k_{\rm mu}$ is given by Eq.~(\ref{kmu}), and 
\begin{eqnarray}
\label{tau1}
\tau & = & \frac{\hbar}{q_{\rm c}}, \\
\xi & = & \sqrt{\frac{8 \hbar^2}{M} \frac{q_{\rm c} - 2 q}{q_{\rm c}^2}}.
\label{xi1}
\end{eqnarray}
For $q_{\rm c} / 2 < q < q_{\rm c}$, $k_{\rm mu} = 0$ and
\begin{eqnarray}
\label{tau2}
\tau & = & \frac{\hbar}{2 \sqrt{q (q_{\rm c} - q)}}, \\
\label{xi2}
\xi & = & \sqrt{\frac{\hbar^2}{M} \frac{2 q - q_{\rm c}}{q (q_{\rm c} -
q)}}.
\end{eqnarray}
At $q = q_{\rm c} / 2$, Eqs.~(\ref{xi1}) and (\ref{xi2}) vanish, and the
$\Delta k^4$ term in Eq.~(\ref{Ekexpand}) becomes important, with
\begin{equation} \label{Xi}
\Xi = 4 \left( \frac{\hbar^4}{2 M^2 q_{\rm c}^2} \right)^{1/4}.
\end{equation}

We first consider a 1D system with the periodic boundary condition, i.e.,
the 1D ring geometry.
We assume that the radius of the ring $R$ is much larger than the domain
size, and the curvature of the ring does not affect the dynamics.

For $q < q_{\rm c} / 2$, the magnetic correlation function is calculated
to be
\begin{eqnarray} \label{ff1}
\left\langle \hat F_+(\theta, t) \hat F_-(\theta', t) \right\rangle & = &
\frac{2 n}{\xi} \sqrt{\frac{\tau}{\pi t}} \cos [k_{\rm mu} R(\theta -
\theta')] \nonumber \\
& & \times e^{t / \tau - \tau R^2 (\theta - \theta')^2 / (t \xi^2)},
\end{eqnarray}
where $\tau$ and $\xi$ are given by Eqs.~(\ref{tau1}) and (\ref{xi1}), and
$\theta$ and $\theta'$ are azimuthal angles.
For $q_{\rm c} / 2 < q < q_{\rm c}$, we obtain
\begin{equation} \label{ff0}
\left\langle \hat F_+(\theta, t) \hat F_-(\theta', t) \right\rangle =
\frac{n}{2 \xi} \sqrt{\frac{\tau}{\pi t}} \frac{q_{\rm c}}{q_{\rm c} - q}
e^{t / \tau - \tau R^2 (\theta - \theta')^2 / (t \xi^2)}
\end{equation}
with Eqs.~(\ref{tau2}) and (\ref{xi2}).
At $q = q_{\rm c} / 2$, the correlation function reads
\begin{eqnarray}
& & \left\langle \hat F_+(\theta, t) \hat F_-(\theta', t) \right\rangle =
\nonumber \\
&  &
\frac{n}{2 \pi \Xi} \frac{q_{\rm c}}{q_{\rm c} - q} \left( \frac{\tau}{t}
\right)^{1/4} e^{t / \tau}
\Biggl[ \Gamma \! \left( \frac{1}{4} \right) {}_0F_2 \! \left(
\frac{1}{2}, \frac{3}{4}, \frac{\tau R^4 (\theta - \theta')^4}{t \Xi^4}
\right)
\nonumber \\
& & - 8 \sqrt{\frac{\tau}{t}} \frac{R^2 (\theta - \theta')^2}{\Xi^2}
\Gamma \! \left( \frac{3}{4} \right) {}_0F_2 \! \left( \frac{5}{4},
\frac{3}{2}, \frac{\tau R^4 (\theta - \theta')^4}{t \Xi^4} \right) \Biggr],
\label{ff3}
\end{eqnarray}
where $\Gamma$ is the Gamma function and
\begin{equation}
{}_0F_2(a, b, z) = \sum_{j = 0}^\infty \frac{\Gamma(a) \Gamma(b)}{\Gamma(a
+ j) \Gamma(b + j)} \frac{z^n}{j!}
\end{equation}
is the generalized hypergeometric function.
Equation~(\ref{ff3}) is shown in Fig.~\ref{f:r-dep}(a), where $\Xi$ gives
a characteristic width of the correlation function.

Next, we consider the 2D geometry.
For $q_{\rm c} / 2 < q < q_{\rm c}$, and then $k_{\rm mu} = 0$, the
integral can be performed analytically, giving
\begin{equation}
\left\langle \hat F_+(\bm{r}, t) \hat F_-(\bm{r}', t) \right\rangle
= \frac{n \tau}{2 \pi \xi^2 t} \frac{q_{\rm c}}{q_{\rm c} - q}
e^{t / \tau - \tau |\bm{r} - \bm{r}'|^2 / (t \xi^2)},
\end{equation}
where $\tau$ and $\xi$ are given in Eqs.~(\ref{tau2}) and (\ref{xi2}).
For other $q$, we can perform only the angular integral as
\begin{eqnarray} \label{ff2d}
\left\langle \hat F_+(\bm{r}, t) \hat F_-(\bm{r}', t) \right\rangle
& = & \frac{n}{4\pi} \frac{q_{\rm c}}{q_{\rm c} - q - \varepsilon_{\rm mu}}
\nonumber \\
& & \times \int_0^\infty k J_0(k |\bm{r} - \bm{r}'|) e^{2 |E_k| t / \hbar}
dk,
\nonumber \\
\end{eqnarray}
where $J_0$ is the Bessel function.
If the exponential factor is much sharper than the Bessel function around
$k_{\rm mu}$, the correlation function (\ref{ff2d}) is approximated to be
$\propto J_0(k_{\rm mu} |\bm{r} - \bm{r}'| )$~\cite{Lamacraft,Uhlmann}.

As shown above, the correlation function (\ref{ff}) has a finite
correlation length, and the magnetization at positions widely separated
from each other grow with independent directions in the $x$-$y$ plane.
Thus, the growth of the magnetic domains is expected to leave topological
defects through the KZ mechanism.

\subsection{Slow quench}

In the previous sections, we have assumed that the magnetic field is
suddenly quenched to the desired value at $t = 0$ and $q$ is held constant
for $t > 0$.
We assume here that for $t > 0$ the magnetic field is gradually quenched
as
\begin{equation} \label{qt}
q(t) = q_{\rm c} \left( 1 - \frac{t}{\tau_{\rm Q}} \right).
\end{equation}

The magnetic correlation can be estimated to be
\begin{eqnarray}
& & \left\langle \hat F_+(\bm{r}, t) \hat F_-(\bm{r}', t) \right\rangle
\nonumber \\
& \propto & \int d\bm{k} \exp \left[\int \frac{2 |E_k(t)| t}{\hbar} dt +
i \bm{k} \cdot (\bm{r} - \bm{r}') \right].
\end{eqnarray}
Since we are interested in the vicinity of the critical point where
correlation starts to grow, we expand $|E_k(t)|$ around $k_{\rm mu} = 0$
and keep the terms up to the order of $k^2$.
For the 1D ring, we obtain
\begin{equation}
\left\langle \hat F_+(\theta, t) \hat F_-(\theta', t) \right\rangle
\propto e^{f(t) - R^2 (\theta - \theta')^2 / \xi_{\rm Q}^2},
\end{equation}
and for the 2D geometry,
\begin{equation}
\left\langle \hat F_+(\bm{r}, t) \hat F_-(\bm{r}', t) \right\rangle
\propto e^{f(t) - |\bm{r} - \bm{r}'|^2 / \xi_{\rm Q}^2},
\end{equation}
where
\begin{eqnarray}
f(t) & = & \frac{\tau_{\rm Q} q_{\rm c}}{2 \hbar} \Biggl[ \tan^{-1}
\sqrt{\frac{t}{\tau_{\rm Q} - t}}
\nonumber \\
& & - \sqrt{\frac{t}{\tau_{\rm Q}} \left(1 -
\frac{t}{\tau_{\rm Q}} \right)} \left( 1 - \frac{2t}{\tau_{\rm Q}}\right)
\Biggr],
\\
\xi_{\rm Q} & = & \left[ \frac{4 \hbar}{M} \sqrt{t (\tau_{\rm Q} - t)}
\right]^{1/2}.
\label{xiQ}
\end{eqnarray}
For $t \ll \tau_{\rm Q}$, $f(t)$ can be expanded as
\begin{equation}
f(t) = \frac{\tau_{\rm Q} q_{\rm c}}{2 \hbar} \left[ \frac{8}{3}
\frac{t^{3/2}}{\tau_{\rm Q}^{3/2}} + O \left( \frac{t^{5/2}}{\tau_{\rm
Q}^{5/2}} \right) \right],
\end{equation}
and from $f(t) \sim 1$, the time scale for magnetization is given by
\begin{equation} \label{tkz}
t_{\rm Q} \sim \left( \frac{\hbar}{q_{\rm c}} \right)^{2/3} \tau_{\rm
Q}^{1/3}.
\end{equation}
Substitution of $t_{\rm Q}$ into Eq.~(\ref{xiQ}) yields
\begin{equation} \label{xikz}
\xi_{\rm Q} \sim \left( \frac{\hbar^4}{M^3 q_{\rm c}} \right)^{1/6}
\tau_{\rm Q}^{1/3}.
\end{equation}
The same power law is obtained in Ref.~\cite{Lamacraft}.

It is interesting to note that the results (\ref{tkz}) and (\ref{xikz})
are easily obtained also by the simple discussion by Zurek~\cite{Zurek}.
Since $q(t)$ depends on time, $\tau$ and $\xi$ given in Eqs.~(\ref{tau2})
and (\ref{xi2}) are time dependent, and hence they are regarded as the
growth time and correlation length at each instant of time.
The local magnetization is developed after a time $t_{\rm Q}$ has elapsed
such that
\begin{equation} \label{zurek}
\tau(t_{\rm Q}) \sim t_{\rm Q}.
\end{equation}
Using
\begin{equation}
\tau(t) = \frac{\hbar \tau_{\rm Q}}{2 q_{\rm c} \sqrt{t (\tau_{\rm Q} - t)}}
\simeq \frac{\hbar \sqrt{\tau_{\rm Q}}}{2 q_{\rm c} \sqrt{t}},
\end{equation}
we obtain $t_{\rm Q}$ in Eq.~(\ref{tkz}).
Substituting this $t_{\rm Q}$ into
\begin{equation}
\xi^2(t) = \frac{\hbar^2}{M q_{\rm c}} \frac{\tau_{\rm Q} - 2 t}{t
(\tau_{\rm Q} - t)} \simeq \frac{\hbar^2
\tau_{\rm Q}}{M q_{\rm c} t}
\end{equation}
yields Eq.~(\ref{xikz}).

\section{Numerical Simulations and the Kibble-Zurek mechanism}
\label{s:num}

\subsection{Gross-Pitaevskii equation with quantum fluctuations}

The multicomponent Gross-Pitaevskii (GP) equation is obtained by replacing
the field operators $\hat \psi_m$ with the macroscopic wave function
$\psi_m$ in the Heisenberg equation of motion:
\begin{subequations}
\label{GP}
\begin{eqnarray}
i \hbar \frac{\partial \psi_{\pm 1}}{\partial t} & = & \left(
-\frac{\hbar^2}{2 M} \nabla^2 + V_{\rm trap} + q + c_0 \rho \right)
\psi_{\pm 1}
\nonumber \\
& & + c_1 \left( \frac{1}{\sqrt{2}} F_\mp \psi_0 \pm F_z
\psi_{\pm 1} \right),
\label{GP1} \\
i \hbar \frac{\partial \psi_0}{\partial t} & = & \left( -\frac{\hbar^2}{2
M} \nabla^2 + V_{\rm trap} + c_0 \rho \right) \psi_0
\nonumber \\
& & + \frac{c_1}{\sqrt{2}} \left(  F_+ \psi_1 + F_- \psi_{-1} \right),
\end{eqnarray}
\end{subequations}
where $\rho$ and $\bm{F}$ are defined using $\psi_m$ instead of $\hat
\psi_m$ in Eqs.~(\ref{rho}) and (\ref{F}).
The wave function is normalized as
\begin{equation}
\int d\bm{r} \sum_{m=-1}^1 |\psi_m|^2 = N,
\end{equation}
with $N$ being the number of atoms in the condensate.

Suppose that all atoms are initially in the $m = 0$ state.
It follows then from Eq.~(\ref{GP1}) that $\psi_{\pm 1}$ will remain zero
in the subsequent time evolution.
This is because quantum fluctuations in the transverse magnetization that
trigger the growth of magnetization are neglected in the mean-field
approximation.
We therefore introduce an appropriate initial noise in $\psi_{\pm 1}$ so
that the mean-field approximation reproduces the quantum evolution.

Let us write the initial state as
\begin{equation} \label{psiini}
\psi_{\pm 1}(\bm{r}) = \sum_{\bm{k}} \frac{1}{\sqrt{V}} e^{i \bm{k} \cdot
\bm{r}} a_{\pm 1, \bm{k}}(0),
\end{equation}
where $a_{\pm 1, \bm{k}}$ are c-numbers.
We assume that the c-number amplitudes $a_{\pm 1, \bm{k}}(0)$ are
stochastic variables whose average values vanish,
\begin{equation} \label{avg}
\langle a_{\pm 1, \bm{k}}(0) \rangle_{\rm avg} = 0,
\end{equation}
where by $\langle \cdots \rangle_{\rm avg}$ we denote the statistical
average over an appropriate probability distribution.
The linear approximation of the GP equation with respect to $a_{\pm 1,
\bm{k}}$ gives the same time evolution as Eq.~(\ref{at}), in which the
operators are replaced by the c-numbers.
We thus obtain
\begin{eqnarray} 
F_+(\bm{r}, t) F_-(\bm{r}', t) & = & \frac{2 n}{V} \sum_{\bm{k}} \left| \cos
\frac{E_k t}{\hbar} + i \frac{\varepsilon_k + q}{E_k} \sin \frac{E_k
t}{\hbar} \right|^2
\nonumber \\
& & \times \Big[ e^{-i \bm{k} \cdot (\bm{r} - \bm{r}')} |a_{1,
\bm{k}}(0)|^2 
\nonumber \\
& & + e^{i \bm{k} \cdot (\bm{r} - \bm{r}')} |a_{-1,
-\bm{k}}(0)|^2 \Bigr].
\label{ffgp}
\end{eqnarray}
Comparing Eq.~(\ref{ffgp}) with Eq.~(\ref{ffa}), we find that they have
the same form if the variance of $a_{\pm 1, \bm{k}}(0)$ satisfies
\begin{equation} \label{var}
\langle |a_{\pm 1, \bm{k}}(0)|^2 \rangle_{\rm avg} = \frac{1}{2}
\end{equation}
for all $\bm{k}$.

In the following, we will perform numerical simulation of spontaneous
magnetization using the GP equation and show that the ensuing dynamics
exhibits defect formation similar to the KZ mechanism.
As the initial state of the $m = \pm 1$ wave functions, we use
Eq.~(\ref{psiini}) with
\begin{equation} \label{rnd}
a_{\pm 1, \bm{k}}(0) = \alpha_{\rm rnd} + i \beta_{\rm rnd},
\end{equation}
where $\alpha_{\rm rnd}$ and $\beta_{\rm rnd}$ are random variables
following the normal distribution $p(x) = \sqrt{2 / \pi} \exp(-2 x^2)$.
Equation (\ref{rnd}) then satisfies Eqs.~(\ref{avg}) and (\ref{var}).

\subsection{1D ring geometry}

Let us first investigate the 1D ring system.
Experimentally this geometry can be realized, e.g., by an optical trap
using a Laguerre-Gaussian beam~\cite{Kuga}.
We reduce the GP equation (\ref{GP}) to 1D by assuming that the wave
function $\psi_m$ depends only on the azimuthal angle $\theta$.
The average density of atoms is assumed to be $n = 2.8 \times 10^{14}$
${\rm cm}^{-3}$.
When the radius of the ring is $R = 50$ $\mu{\rm m}$ and the radius of the
small circle is 2 $\mu{\rm m}$, the total number of atoms is $N \simeq
10^6$.

\begin{figure}[t]
\includegraphics[width=9cm]{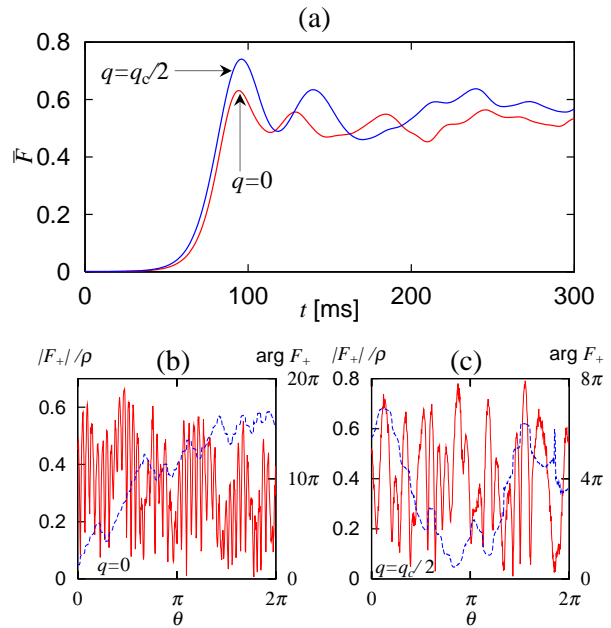}
\caption{
(Color online) (a) Time evolution of the auto correlation function given
in Eq.~(\ref{autoc}) for the 1D ring geometry.
(b) Magnitude of the normalized magnetization $|F_+| / \rho$ (solid curve,
left axis) and direction of the magnetization ${\rm arg} F_+$ (dashed
curve, right axis) at $t = 70$ ms for $q = 0$ and (c) for $q = q_{\rm c} /
2$.
The radius of the ring is $R = 50$ $\mu{\rm m}$, the atomic density is $n
= 2.8 \times 10^{14}$ ${\rm cm}^{-3}$, and the number of atoms is $N =
10^6$.
}
\label{f:1dev}
\end{figure}
Figure~\ref{f:1dev} illustrates a single run of time evolution for an
initial state given by Eqs.~(\ref{psiini}) and (\ref{rnd}).
Figure~\ref{f:1dev} (a) shows time evolution of the auto correlation
function defined by
\begin{equation} \label{autoc}
\bar{F}(t) = \int R d\theta \frac{|F_+(\theta, t)|^2}{\rho^2(\theta, t)}.
\end{equation}
For both $q = 0$ and $q = q_{\rm c} / 2$, the transverse magnetization
grows exponentially with a time constant $\sim \tau = \hbar / q_{\rm c}
\simeq 8$ ms.
Snapshots of the transverse magnetization at $t = 70$ ms are shown in
Figs.~\ref{f:1dev} (b) and \ref{f:1dev} (c) for $q = 0$ and $q = q_{\rm c}
/ 2$, respectively.
We define the spin winding number as
\begin{equation}
w = \frac{1}{2\pi} \int_0^{2\pi} R d\theta \frac{1}{2i |F_+|^2}
\left( F_- \frac{\partial F_+}{\partial \theta} - F_+ \frac{\partial
F_-}{\partial \theta} \right),
\end{equation}
which represents the number of rotation of the spin vector in the $x$-$y$
plane along the circumference, and of course $w$ is an integer.
The spin winding numbers are $w = 7$ in Fig.~\ref{f:1dev} (b) and $w = -1$
in Fig.~\ref{f:1dev} (c).

\begin{figure}[t]
\includegraphics[width=9cm]{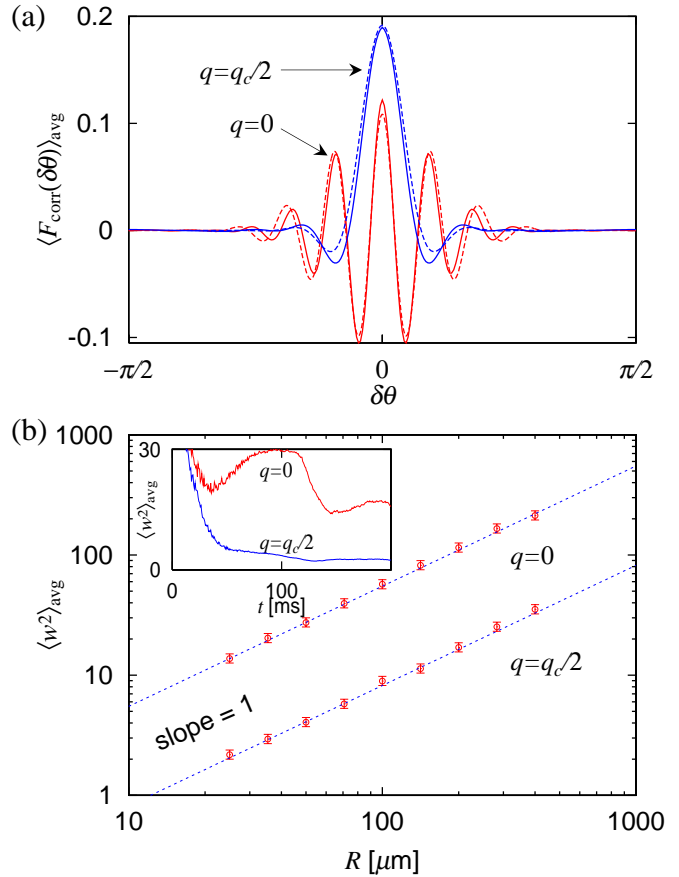}
\caption{
(Color online) (a) Numerically obtained correlation
function given in Eq.~(\ref{fcorr}) at $t = 70$ ms (solid curves), and
theoretical fits (dashed curves) from Eqs.~(\ref{ff1}) and (\ref{ff3}).
Other parameters are the same as those in Fig.~\ref{f:1dev}.
(b) $R$ dependence of the variance of the spin winding number, where the
number of atoms is related to $R$ as $N = 10^6 \times R$ $[\mu{\rm m}]$ $/
50$.
The dashed lines are semi-log fits to the numerical data.
The inset shows the time dependence of $\langle w^2 \rangle_{\rm avg}$ for
$R = 50$ $\mu{\rm m}$.
The data in (a) and (b) are averages over 1000 runs of simulations for
different initial states produced by random numbers.
The error bars in (b) represent the 95\% confidence interval of the
$\chi^2$ distribution.
}
\label{f:r-dep}
\end{figure}
Figure~\ref{f:r-dep} (a) shows the ensemble average of the normalized
correlation function,
\begin{equation} \label{fcorr}
\langle F_{\rm corr}(\delta \theta) \rangle_{\rm avg} = \left\langle
\frac{\int d\theta F_+(\theta) F_-(\theta + \delta\theta)}{\int d\theta
\rho(\theta) \rho(\theta + \delta\theta)} \right\rangle_{\rm avg},
\end{equation}
at $t = 70$ ms.
For $q = q_{\rm c} / 2$, the correlation function has the characteristic
width of $\sim \Xi$ in Eq.~(\ref{Xi}), indicating that the ring is filled
with magnetic domains with an average size of $\sim \Xi$.
According to the KZ theory, the magnetic domains with random directions
give rise to the spin winding, which is estimated to be $w \sim (R /
\Xi)^{1/2}$.
This $R$ dependence of $w$ is clearly seen in Fig.~\ref{f:r-dep} (b).
The ensemble average of the winding number, $\langle w \rangle_{\rm avg}$,
vanishes due to the random nature of the initial noise, and the square
root of its variance, $\langle w^2 \rangle_{\rm avg}^{1/2}$, should be
regarded as a typical winding number.
The variance is expected to obey the $\chi^2$ distribution with 1000
degrees of freedom, and hence we show the 95\% confidence interval to
estimate the statistical errors in Fig.~\ref{f:r-dep}.
As shown in the inset of Fig.~\ref{f:r-dep} (b), the typical winding
number changes in time, since the ferromagnetic energy is converted to the
kinetic energy and the system exhibits complicated dynamics.

The situation is different for $q = 0$, in which the correlation function
oscillates with a Gaussian envelope as shown in Fig.~\ref{f:r-dep} (a).
This form of the correlation function gives us the answer to the question
as to how the KZ mechanism manifests itself in spin conserving systems.
The finite correlation length for $q = 0$ indicates that the spin is
conserved not only globally but also locally, that is, the locally
integrated spin over the correlation length $\xi$,
\begin{equation}
\int_{|\delta\bm{r}| \lesssim \xi} \bm{F}(\bm{r} + \delta\bm{r})
d\delta\bm{r},
\end{equation}
is held to be zero for any $\bm{r}$.
This local spin conservation is due to formation of staggered domain or
helical spin structures whose periodic length is much smaller than $\xi$.
Thus, the neighboring domains tend to have opposite magnetizations to
cancel out the spin locally, and the domains far from each other grow
independently; the spin conservation and the KZ mechanism are thus
compatible.

\begin{figure}[t]
\includegraphics[width=9cm]{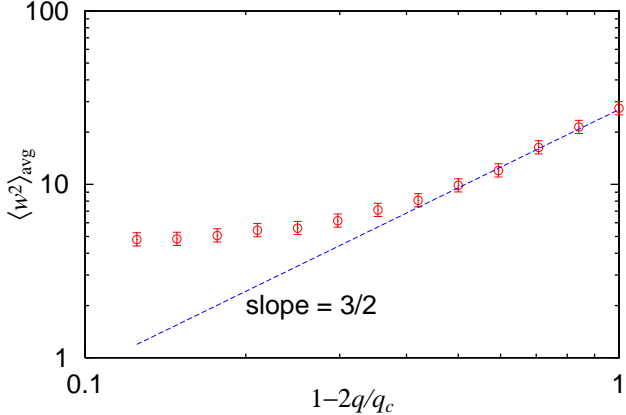}
\caption{
(Color online) 
Dependence of the variance of the spin winding number on $q$.
Except for $q$, the parameters are the same as those in
Fig.~\ref{f:1dev}.
The dashed line is proportional to $(1 - 2 q / q_{\rm c})^{3/2}$.
The plots show the averages over 1000 runs of simulations for different
initial states produced by random numbers.
The error bars represent the 95\% confidence interval of the $\chi^2$
distribution.
}
\label{f:kmax-dep}
\end{figure}
The oscillation in the correlation function originates from the fact that
the most unstable modes have nonzero wave numbers $\pm k_{\rm mu}$.
Each correlated region of size $\sim \xi = [8 \hbar^2 / (M q_{\rm
c})]^{1/2}$ contains spin waves of $e^{i k_{\rm mu} R \theta}$ and $e^{-i
k_{\rm mu} R \theta}$.
If there is an imbalance between these modes, the winding number
monotonically increases or decreases in each region of $\sim \xi$.
This is the reason why $\langle w^2 \rangle_{\rm avg}$ is larger for $q =
0$ than for $q = q_{\rm c} / 2$ in Fig.~\ref{f:r-dep} (b).
It follows from this consideration that for $k_{\rm mu} \xi \gg 1$ the
winding number is proportional to
\begin{equation} \label{qdep}
w \sim k_{\rm mu} \xi \sqrt{\frac{R}{\xi}} = k_{\rm mu} \sqrt{R \xi}
\propto \left( 1 - \frac{2 q}{q_{\rm c}} \right)^{3/4},
\end{equation}
where Eqs.~(\ref{kmu}) and (\ref{xi1}) are used.
Figure~\ref{f:kmax-dep} shows the averaged variance of the winding number
versus $1 - 2 q / q_{\rm c}$.
For small $q$, $\langle w^2 \rangle_{\rm avg}$ is proportional to $(1 - 2
q / q_{\rm c})^{3/2}$, in agreement with Eq.~(\ref{qdep}).
When $q$ is close to $q_{\rm c} / 2$, the spin winding within the
correlated region, $k_{\rm mu} \xi$, becomes small, and then the winding
number reduces to the value shown in Fig.~\ref{f:r-dep} (b), i.e.,
$\langle w^2 \rangle_{\rm avg} \simeq 4$.

\begin{figure}[t]
\includegraphics[width=9cm]{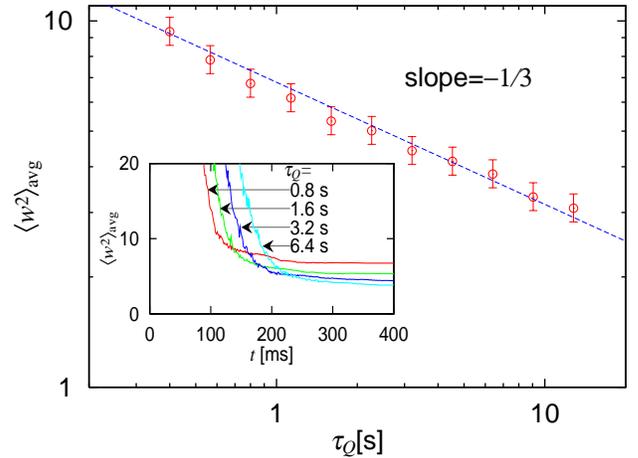}
\caption{
(Color online) 
Dependence of the variance of the spin winding number at $t = 400$ ms on
the quench time $\tau_{\rm Q}$, where $q$ is varied as in Eq.~(\ref{qt}).
The radius of the ring is $R = 400$ $\mu{\rm m}$, the atomic density is $n
= 2.8 \times 10^{14}$ ${\rm cm}^{-3}$, and the number of atoms is $N = 8
\times 10^6$.
The dashed line is proportional to $\tau_{\rm Q}^{-1 / 3}$.
The inset shows time evolution of $\langle w^2 \rangle_{\rm avg}$.
The data are averages over 1000 runs of simulations for different
initial states produced by random numbers.
The error bars represent the 95\% confidence interval of the $\chi^2$
distribution.
}
\label{f:slow}
\end{figure}
We next discuss the results of simulations of slow quench as in
Eq.~(\ref{qt}).
Since the winding number for the slow quench is small compared with the
fast quench, we take a large ring of $R = 400$ $\mu{\rm m}$.
Figure~\ref{f:slow} shows the variance of the winding number as a function
of the quench time.
We can clearly see that $\langle w^2 \rangle_{\rm avg}$ has a power law
of $\tau_{\rm Q}^{-1/3}$ within the statistical error,
which is in agreement with $\xi_{\rm Q}^{-1} \sim \tau_{\rm Q}^{-1/3}$,
with $\xi_{\rm Q}$ being given in Eq.~(\ref{xikz}).
Thus, the present system follow the quench-time scaling of
Zurek~\cite{Zurek}.
We note that, in the slow quench, the winding number converges to an
almost constant value for varying quench time $\tau_{\rm Q}$, as shown in
the inset of Fig.~\ref{f:slow}.
This is because little excess energy other than for exciting spin vortices
is available for the slow quench.

\subsection{2D disk geometry}

When the confinement in the $z$ direction is tight, the system is
effectively 2D.
For simplicity, we ignore the density dependence in the $z$ direction, and
assume that the 2D GP equation has the same form as Eq.~(\ref{GP}).
We assume that the wave function vanishes at the wall located at $(x^2 +
y^2)^{1/2} = R_{\rm w} = 100$ $\mu{\rm m}$, and that the potential is flat
inside of the wall.
Then the density $n = 2.8 \times 10^{14}$ ${\rm cm}^{-3}$ is almost
constant except within the healing length $\{3 / [8 \pi n (a_0 + 2
	a_2)]\}^{1/2} \simeq 0.16$ $\mu{\rm m}$ near the wall.
When the thickness in the $z$ direction is $\simeq 1$ $\mu{\rm m}$, the
number of atoms is $N \simeq 10^7$.
Such a system will be realized using an optical sheet and a hollow laser
beam.

\begin{figure*}[ht]
\includegraphics[width=13.8cm]{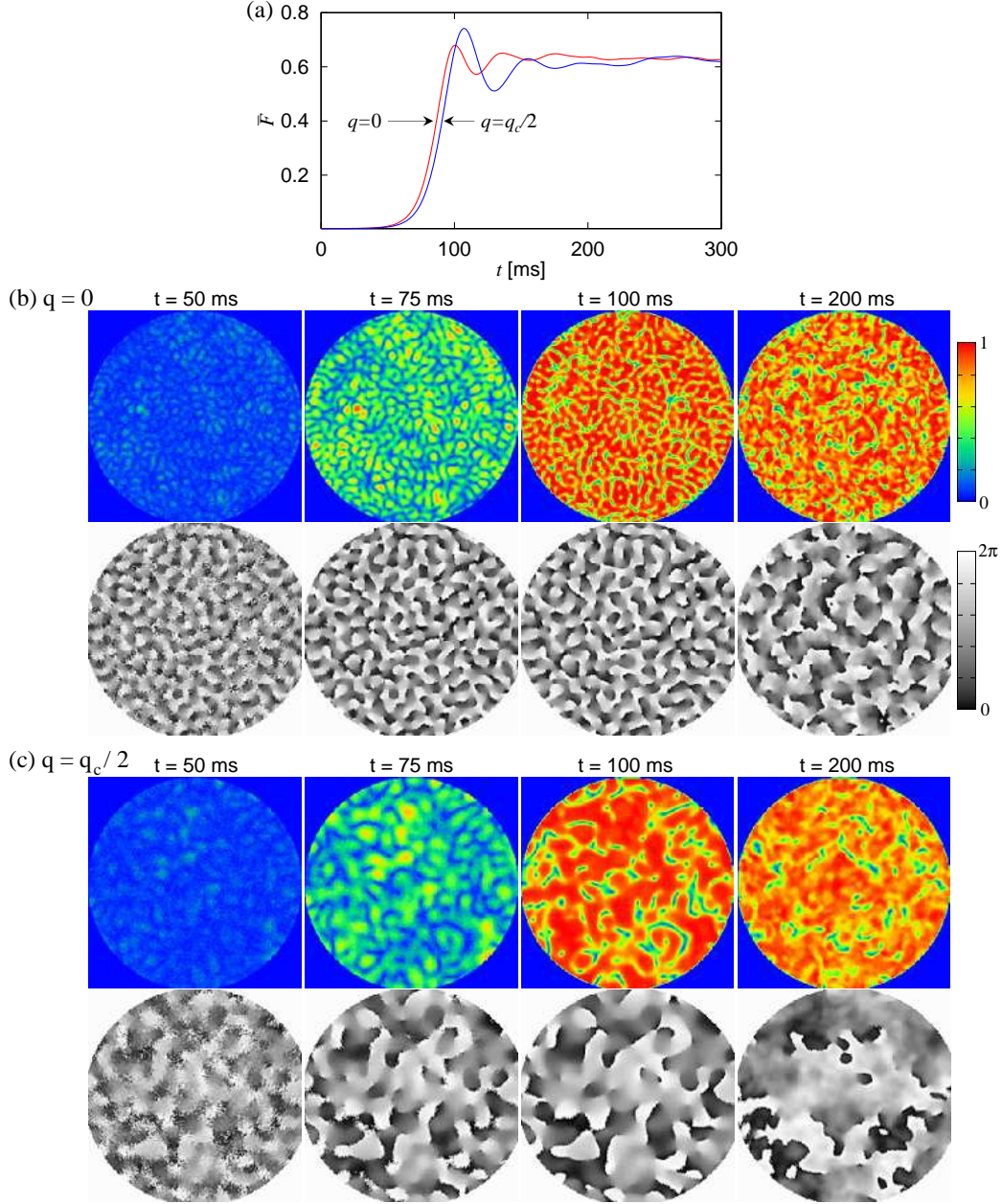}
\caption{
(Color)
(a) Time evolution of the autocorrelation function given in
Eq.~(\ref{autoc2d}) for the 2D disk geometry.
The radius of the disk is $R_{\rm w} = 100$ $\mu{\rm m}$, the atomic
density is $n = 2.8 \times 10^{14}$ ${\rm cm}^{-3}$, and the number of
atoms is $N = 10^7$.
(b) Profiles of the magnetization $|F_+|$ (upper) and its direction ${\rm
arg} F_+$ (lower) for $q = 0$ and (c) for $q = q_{\rm c} / 2$.
The size of each panel is $200$ $\mu {\rm m}$ $\times 200$ $\mu {\rm m}$.
}
\label{f:2d}
\end{figure*}
The initial state of $\psi_0$ is a stationary solution of the GP equation,
and the initial state of $\psi_{\pm 1}$ is given by Eq.~(\ref{psiini})
with random variables (\ref{rnd}).
Figure~\ref{f:2d} (a) shows time evolution of the autocorrelation
function of the transverse magnetization,
\begin{equation} \label{autoc2d}
\bar{F}(t) = \int d\bm{r} \frac{|F_+(\bm{r}, t)|^2}{\rho^2(\bm{r}, t)},
\end{equation}
which grows exponentially with the same time constant as that in
Fig.~\ref{f:1dev}, and saturates for $t \gtrsim 100$ ms.

Snapshots of $|F_+(\bm{r})|$ and ${\rm arg} F_+(\bm{r})$ at $t = 100$ ms
are shown in Figs.~\ref{f:2d} (b) and \ref{f:2d} (c).
We see that $|F_+(\bm{r})|$ at $t \gtrsim 100$ ms contains many holes,
around which the spin direction rotates by $2\pi$.
Since this topological spin structure consists of singly-quantized
vortices in the $m = \pm 1$ states filled by atoms in the $m = 0$ state,
it is called the ``polar-core vortex.''
We can estimate the spin healing length $\xi_{\rm s}$ by equating the
kinetic energy $\hbar^2 / (2 M \xi_{\rm s}^2)$ with the energy of
magnetization $|q - q_{\rm c}|$, giving
\begin{equation}
\xi_{\rm s} = \frac{\hbar}{\sqrt{2 M |q - q_{\rm c}|}}.
\end{equation}
This length scale is $\xi_{\rm s} \simeq 1.7$ $\mu{\rm m}$ for $q = 0$ and
$\xi_{\rm s} \simeq 2.4$ $\mu{\rm m}$ for $q = q_{\rm c} / 2$, which are
in good agreement with the sizes of the vortex cores in Figs.~\ref{f:2d}
(b) and \ref{f:2d} (c).

\begin{figure}[t]
\includegraphics[width=9cm]{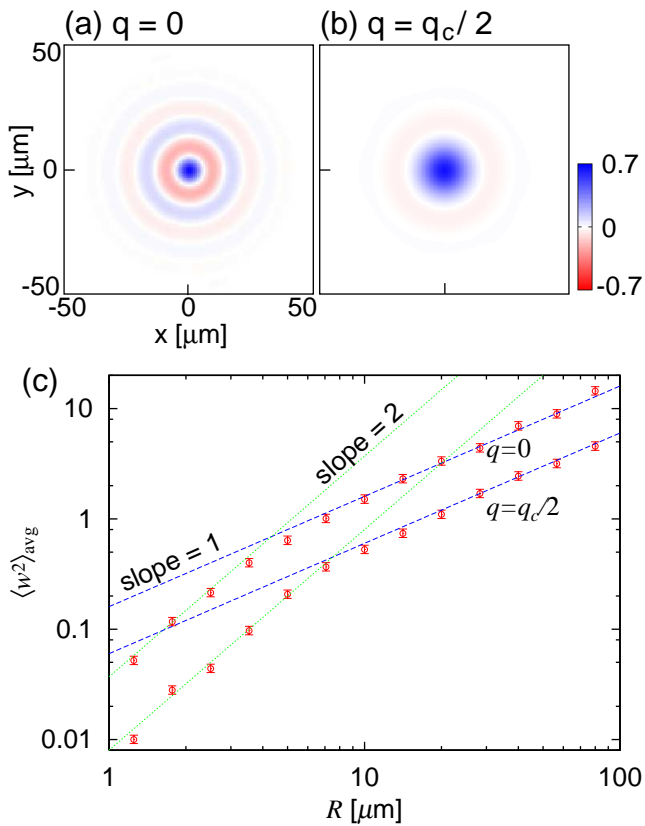}
\caption{
(Color)
(a) Spin correlation function defined in Eq.~(\ref{fcorr2d}) at $t = 100$
ms for $q = 0$ and (b) for $q = q_{\rm c} / 2$.
(c) The variance of the winding number along the circumference of the
circle of radius $R$.
The dashed lines and dotted lines are proportional to $R$ and $R^2$,
respectively.
In (a)-(c) the parameters are the same as those in Fig.~\ref{f:2d}, and
the data are averages over 1000 runs of simulations for different initial
states produced by random numbers.
The error bars in (c) represent the 95\% confidence interval of the
$\chi^2$ distribution.
}
\label{f:2drdep}
\end{figure}
In 2D, the correlation function is defined by
\begin{equation} \label{fcorr2d}
\langle F_{\rm corr}(\delta \bm{r}) \rangle_{\rm avg} = \left\langle
\frac{\int d\bm{r} F_+(\bm{r}) F_-(\bm{r} + \delta\bm{r})}{\int d\bm{r}
\rho(\bm{r}) \rho(\bm{r} + \delta\bm{r})} \right\rangle_{\rm avg},
\end{equation}
which are shown in Figs.~\ref{f:2drdep} (a) and \ref{f:2drdep} (b).
We find that as in the 1D case the most unstable wave length is reflected
in the shape of the spin correlation function~(\ref{fcorr2d}), and the
characteristics of these correlation functions in the radial direction are
similar to those in 1D shown in Fig.~\ref{f:r-dep}.
For $q = 0$, the mean distance between spin vortices in Fig.~\ref{f:2d}
(b) is not determined by the correlation length (the whole width of the
concentric pattern in Fig.~\ref{f:2drdep} (a)) but by $\sim k_{\rm
mu}^{-1}$, i.e., the width of the concentric rings in Fig.~\ref{f:2drdep}
(a).
On the other hand, for $q = q_{\rm c} / 2$, the density of spin vortices
is determined by the correlation length, i.e., the size of the blue circle
$\simeq 30$ $\mu {\rm m}$ in Fig.~\ref{f:2drdep} (b).
The staggered concentric correlation for $q = 0$ suggests that the spin is
conserved locally within the region of the correlation length, and domains
at a distance larger than the correlation length grow independently, while
conserving the total spin.

The spin winding number for 2D is defined as
\begin{equation}
w(R) = \frac{1}{2\pi} \oint_{C(R)} \frac{1}{2i |F_+|^2} \left( F_-
\bm{\nabla} F_+ - F_+ \bm{\nabla} F_- \right) \cdot d\bm{r},
\end{equation}
where $C(R)$ is a circle with radius $R < R_{\rm w}$ located at the center
of the system.
Figure~\ref{f:2drdep} (c) shows the $R$ dependence of the ensemble average
of $w^2(R)$, where the radius of the system is fixed to $R_{\rm w} = 100$
$\mu{\rm m}$ and the data are taken at $t = 100$ ms.
It should be noted that $\langle w^2(R) \rangle_{\rm avg}$ is proportional
to $R$ for large $R$, as expected from the KZ theory~\cite{Zurek}, while
it is proportional to $R^2$ for small $R$.
This $R^2$ dependence is due to the fact that the probability $P$ for a
spin vortex to be in the circle is proportional to $\pi R^2$.
The variance of the winding number is therefore $0 (1 - P) + 1^2 P / 2 +
(-1)^2 P / 2 \propto R^2$, if the probability that two or more vortices
enter the circle is negligible.
This condition is met when the density of spin vortices times $\pi R^2$
is much smaller than unity, and hence the radius $R$ at which the
crossover from $\langle w^2(R) \rangle_{\rm avg} \propto R$ to $\propto
R^2$ occurs is larger for $q = q_{\rm c} / 2$ than for $q = 0$.
As in 1D, nonzero $k_{\rm mu}$ enhances the winding of magnetization, and
the winding number is larger for $q = 0$ than for $q = q_{\rm c} / 2$.

Figures~\ref{f:2d} (b) and \ref{f:2d} (c) obviously show that the density
of spin vortices is uniform when the size of the system is large enough.
The number of spin vortices in a radius $R$ is therefore proportional to
$R^2$.
If the topological charge of each spin vortex, $+1$ or $-1$, was chosen at
random, the net winding number along the circle of radius $R$, i.e., the
difference between the numbers of $+1$ and $-1$ vortices would be
proportional to $R$.
However, from Fig.~\ref{f:2drdep} (c), the winding number is proportional
to $R^{1/2}$ for large $R$, consistent with the KZ mechanism.
The topological charge of each spin vortex is thus not at random but
anticorrelated to each other to reduce the net winding number.

\begin{figure}[t]
\includegraphics[width=9cm]{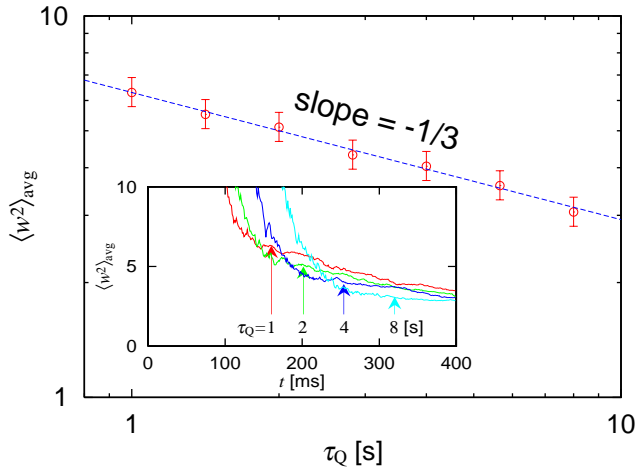}
\caption{
(Color online)
(a) Variance of the spin winding number versus the quench time $\tau_{\rm
Q}$ for the 2D disk geometry, where $q$ is varied as in Eq.~(\ref{qt}).
The inset shows time evolution of $\langle w^2 \rangle_{\rm avg}$.
The plots are taken at the times when $t / \tau_{\rm Q}^{1/3} =$ constant
is satisfied, which are shown by the arrows in the inset.
The dashed line is proportional to $\tau_{\rm Q}^{-1/3}$.
The radius of the disk is $R_{\rm w} = 400$ $\mu {\rm m}$ and the closed
path for taking the winding number is $R = 320$ $\mu {\rm m}$.
The atomic density is $n = 2.8 \times 10^{14}$ ${\rm cm}^{-3}$ and the
number of atoms is $N = 1.6 \times 10^8$.
The data are averages over 1000 runs of simulations for different initial
states produced by random numbers.
The error bars represent the 95\% confidence interval of the $\chi^2$
distribution.
}
\label{f:2dslow}
\end{figure}
Figure~\ref{f:2dslow} shows the result of the slow quench for 2D, where
$q(t)$ is given by Eq.~(\ref{qt}).
The winding number follows the scaling law, $\langle w^2 \rangle_{\rm avg}
\propto \tau_{\rm Q}^{-1/3}$, as predicted from Eq.~(\ref{xikz}),
indicating that Zurek's discussion is applicable also to 2D.
In order to obtain this scaling law, we must specify the time at which the
winding number is taken, since the spin winding number decays in time, as
shown in the inset of Fig.~\ref{f:2dslow}.
From the scaling law in Eq.~(\ref{tkz}), we specify the time to take the
winding number as
\begin{equation}
\frac{t}{\tau_{\rm Q}^{1/3}} \left( \frac{q_{\rm c}}{\hbar} \right)^{2/3}
= {\rm const.},
\end{equation}
which is indicated by the arrows in the inset of Fig.~\ref{f:2dslow}.

\section{Conclusions}
\label{s:conc}

In this paper, we have studied the dynamics of a spin-1 BEC with a
ferromagnetic interaction after quench of the applied magnetic field in an
attempt to investigate spontaneous defect formation in the spinor BEC.
We have analyzed the magnetization triggered by quantum fluctuations using
the Bogoliubov approximation, and performed numerical simulations of the
GP equation with initial conditions that simulate quantum fluctuations.

We have shown that the correlation functions of the magnetization have
finite correlation lengths (Figs.~\ref{f:r-dep}, \ref{f:2drdep} (a), and
\ref{f:2drdep} (b)), and therefore magnetic domains far from each other
grow in random directions.
We find that topological defects --- spin vortices --- emerge through the
KZ mechanism.
We have confirmed that the winding number along the closed path is
proportional to the square root of the length of the path
(Figs.~\ref{f:r-dep} (b) and \ref{f:2drdep} (c)), indicating that the
topological defects are formed from domains with random directions of
magnetizations.

Even when the total magnetization is conserved for $q = 0$, the winding
number has the same dependence on the length of the path
(Fig.~\ref{f:r-dep} (b)).
This is due to the fact that domains within the correlation length tend to
be aligned in such a manner as to cancel out local magnetization, and
consequently the total magnetization is conserved.
Thus, the neighboring domains have local correlation, while domains far
from each other are independent, which makes the KZ mechanism compatible
with the total spin conservation.
The formation of the local correlation also creates topological defects as
well as the KZ mechanism, and the winding number exhibits the $q$
dependence as shown in Fig.~\ref{f:kmax-dep}.

When the magnetic field is quenched in finite time $\tau_{\rm Q}$ as in
Eq.~(\ref{qt}), the winding number has been shown to be proportional to
$\tau_{\rm Q}^{-1/6}$ (Figs.~\ref{f:slow} and \ref{f:2dslow}).
This $\tau_{\rm Q}$ dependence of the winding number can be understood
from Zurek's simple discussion~\cite{Zurek}:
the domains are frozen at which the spin relaxation time becomes the
same order of elapsed time.

In the Berkeley experiment~\cite{Sadler}, the system is an elongated
quasi-2D geometry, and not suitable for testing the KZ mechanism.
The KZ mechanism should apply to the system in which the size of the
system in the $x$ direction is made much larger.
In this case, the harmonic potential may affect the scaling law, which
merits further study.
Moreover in the experiment, from the analysis in Ref.~\cite{Saito07},
there are some initial atoms in the $m = \pm 1$ components with long-range
correlation, which play a role of seeds for large domains and hinder the
observation of the KZ mechanism.
If the residual atoms in the $m = \pm 1$ components is eliminated
completely, magnetization is triggered by quantum fluctuations as shown in
the present paper.
Another way to remove the effect of the residual atoms may be applying
random phases to the $m = \pm 1$ states to erase the initial correlation.

{\it Note added.} After our work was completed, the preprint by Damski and
Zurek~\cite{Damski} appeared, which performs 1D simulations of the quench
dynamics of a spin-1 BEC.

\begin{acknowledgments}
This work was supported by Grants-in-Aid for Scientific Research (Grant
Nos.\ 17740263 and 17071005) and by the 21st Century COE programs on
``Coherent Optical Science'' and ``Nanometer-Scale Quantum Physics'' from
the Ministry of Education, Culture, Sports, Science and Technology of
Japan.
MU acknowledges support by a CREST program of the JST.
\end{acknowledgments}

\end{document}